\documentclass[11pt]{article}

\usepackage{array}
\usepackage{graphicx}
\usepackage{amsmath}
\usepackage{amssymb}
\usepackage{hhline}

\title{\bf Spurious localized highest-frequency modes in 
Schr\"odinger-type equations solved by finite-difference methods}

\author{ T.I. Lakoba\footnote{lakobati@cems.uvm.edu, \ 1 (802) 656-2610} 
 \vspace{0.5cm} \\
  Department of Mathematics and Statistics, 16 Colchester Ave., \\
 University of Vermont, Burlington, VT 05401, USA}

\newcommand{\noi}{\noindent}
\newcommand{\D}{\Delta}
\newcommand{\be}{\begin{equation}}
\newcommand{\ee}{\end{equation}}
\newcommand{\bsube}{\begin{subequations}}
\newcommand{\esube}{\end{subequations}}
\newcommand{\ba}{\begin{array}}
\newcommand{\ea}{\end{array}}

\newcommand{\bea}{\begin{eqnarray}}
\newcommand{\eea}{\end{eqnarray}}

\newcommand{\const}{{\rm const}}
\newcommand{\sech}{\mbox{sech}}

\begin{document}
\baselineskip 18 pt

\maketitle

\vspace*{2cm}

\begin{center}
 {\bf Abstract}
\end{center}

High-frequency solutions of one or several Schr\"odinger-type equations 
are well known to differ very little from the plane wave solutions 
 \ $\exp[\pm ik x]$. That is, the potential terms impact the envelope
of a high-frequency plane wave by only a small amount. 
However, when such equations are solved by a finite-difference method,
the highest-frequency solutions may, under certain conditions, turn
out to be localized. In this letter we explain this numerical artifact.

\vskip 1.1 cm

\noi
{\bf Keywords}: \ Coupled Schr\"odinger equations,
Finite-difference discretization, High-frequency modes.

\bigskip


\newpage


Coupled Schr\"odinger-type equations arise in many areas of physics, e.g., in
quantum mechanics \cite{Fulling75,Johnson78,Teubner06,Skorupski08} and 
stability of nonlinear waves \cite{Chugunova10}.
Often, the problem is posed as an eigenvalue problem whereby localized eigenfunctions 
and their eigenvalues are sought. The eigenfunction localization occurs due to
the presence of potential-like terms. A simple approach of solving such an eigenvalue
problem is to discretize the equations by a finite-difference method and then 
solve the resulting matrix eigenvalue problem by a commercial software. Then, by
inspection or otherwise, one selects the localized eigenfunctions and their eigenvalues
out of a set of eigensolutions produced by the software.

High-frequency solutions are, typically, not sought numerically because
their approximate analytical form can be found by perturbation
methods, e.g., by the Born or Wentzel--Kramers--Brillouin approximations. 
For example, for a single Schr\"odinger equation
\be
-d^2\psi/dx^2 + V(x)\psi = \lambda \psi, \qquad \lambda>0,
\label{e_01}
\ee
where $\lambda\gg 1$ and $V(x)$ varies on the scale of order one and also $\max|V(x)|=O(1)$,
the latter approximation yields:
\be
\psi(x) = \left( \frac{\lambda}{\lambda - V(x)} \right)^{1/4} \;
\exp\left[ \pm i\left\{ \sqrt{\lambda}x-\frac1{\sqrt{\lambda}}\int V(x)\,dx \right\} \, \right]
 \; \left( 1 + O\left( \frac1{\sqrt{\lambda}} \right) \; \right).
\label{e_02}
\ee
Moreover, to accurately resolve a solution with a given frequency, one needs about 
10 grid points per wavelength. The highest-frequency mode
resolved on a grid with a step
size $h$ has a wavelength $2h$; that is, such a mode has only 2 grid points per wavelength.
Thus, one does not expect that such a mode can be resolved with much quantitative accuracy.
However, one does expect that it should qualitatively look like solution \eqref{e_02}:
its envelope is to be a finite constant away from the localized potential $V(x)$ and is to
have a small ``wiggle" around the potential.

The researcher may want to inspect the numerically obtained
highest-frequency solutions of the eigenproblem
in order to verify that at least in the high-frequency limit, his/her finite-difference
code produces reasonable results, as described above. If the code produces {\em qualitatively}
different profiles of high-frequency modes, the researcher may question the correctness of
the code and search for a mistake. This can be a time-consuming task when a system of
several coupled equations is considered. Thus, it is valuable to know what the 
highest-frequency modes obtained by a finite-difference method can look like.

Below we show that such modes look not at all as described at the end of the paragraph
containing Eq.~\eqref{e_02}. Rather, their envelopes are the {\em lowest}-frequency
eigenfunctions of the potential \ $-V(x)$. In short, this occurs because the finite-difference
approximation to $d^2/dx^2$ in \eqref{e_01} evaluated on the highest-frequency
carrier \ $\exp[ikx]$ \ with $k=\pi/h$ \ becomes \ ``$\const - d^2/dx^2$"; 
note the change of sign in front of the second derivative. 
Observing localized envelopes of the numerically obtained highest-frequency modes
may be even more counterintuitive given that they occur for a repulsive rather than
attractive potential.

While we noticed this fact when numerically solving  several coupled Schr\"odinger-type
equations, below we chose to present its explanation for a single equation \eqref{e_01},
so that the complexity of the problem would not obfuscate the essence of the explanation.
As a finite-difference approximation of $d^2\psi/dx^2$ we use the simple central difference:
\be
d^2\psi(x_n)/dx^2 \approx \big(\psi(x_{n+1}) - 2\psi(x_n) + \psi(x_{n-1})\,\big)/h^2,
\label{e_03}
\ee
where $x_{n\pm 1}=x_n\pm h$. Using a more accurate Numerov's discretization \cite{Johnson78}
leads to the same qualitative conclusions.

The following Matlab code computes the four highest-frequency modes and the corresponding
eigenvalues of Eq.~\eqref{e_01} with $V(x)=3\,\sech(0.5x)$ and periodic boundary conditions:
\begin{verbatim}
h=0.1; x=-16:h:16-h; N=length(x);
M=spdiags(repmat([-1 2 -1],N,1),[-1 0 1],N,N)/h^2 + diag(3*sech(0.5*x));
M(1,end)=-1/h^2; M(end,1)=-1/h^2;
[Evecs, Evals]=eigs(M,4,'lm'); absEvecs=abs(Evecs);
k=1; plot(x,sech(0.5*x),'--',x,absEvecs(:,k)/max(absEvecs(:,k)));
\end{verbatim}
The envelopes of the first and fourth such modes are shown in Figs.~1(a,b);
the carrier is shown in Fig.~1(c). As we have announced above, these envelopes are localized,
in contrast to the slightly perturbed plane waves \eqref{e_02} that could have been
expected naively. A calculation that explains Fig.~1 is as follows.

\begin{figure}[h!]
\hspace*{0.3cm}
\begin{minipage}{4.7cm}
 \vspace{-1.2cm}
\rotatebox{0}{\resizebox{4.7cm}{6cm}{\includegraphics[0in,0.5in]
 [8in,10.5in]{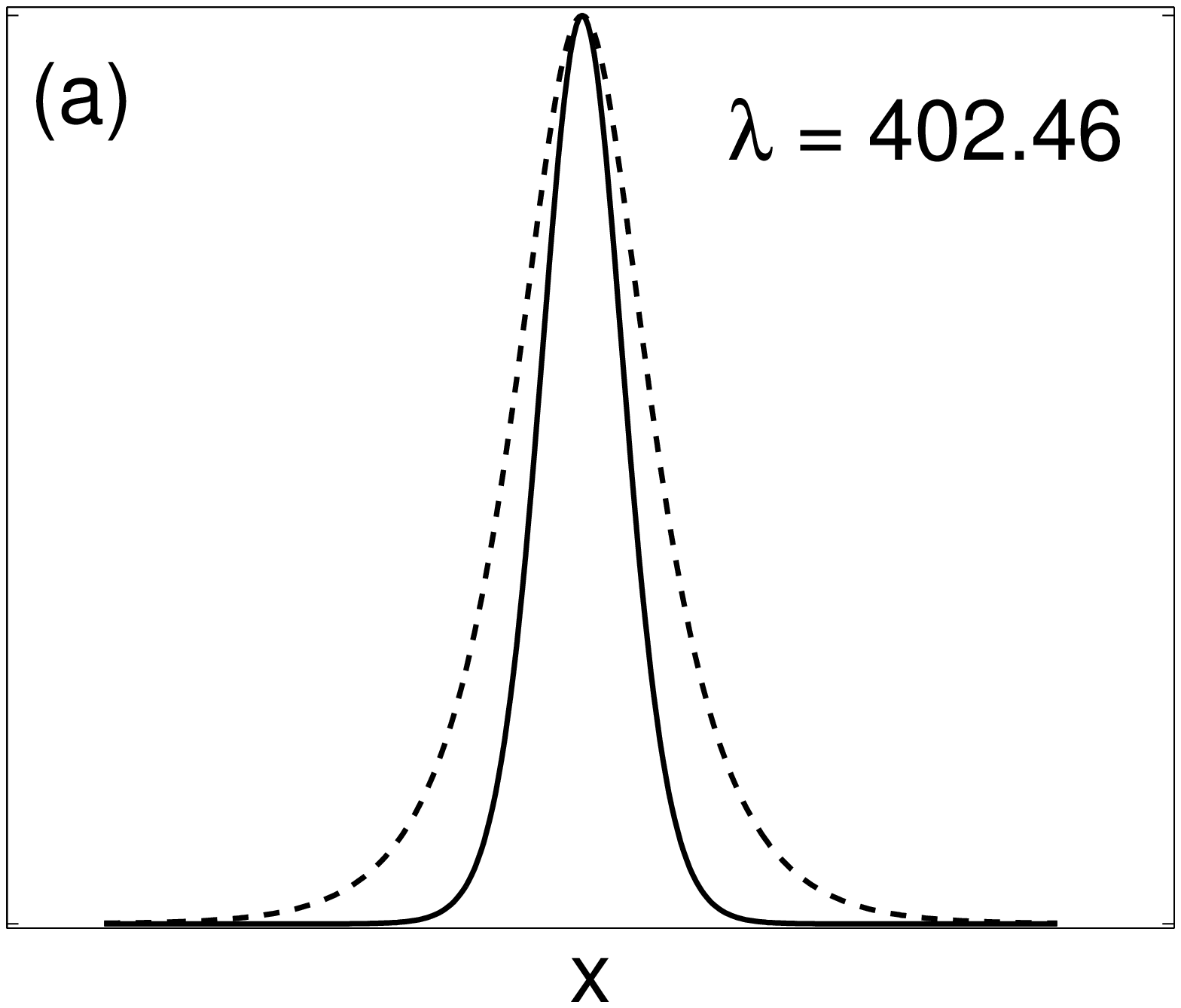}}}
 \end{minipage}
 \hspace{-0.1cm}
 \begin{minipage}{4.7cm}
 \vspace{-1.2cm}
 \rotatebox{0}{\resizebox{4.7cm}{6cm}{\includegraphics[0in,0.5in]
  [8in,10.5in]{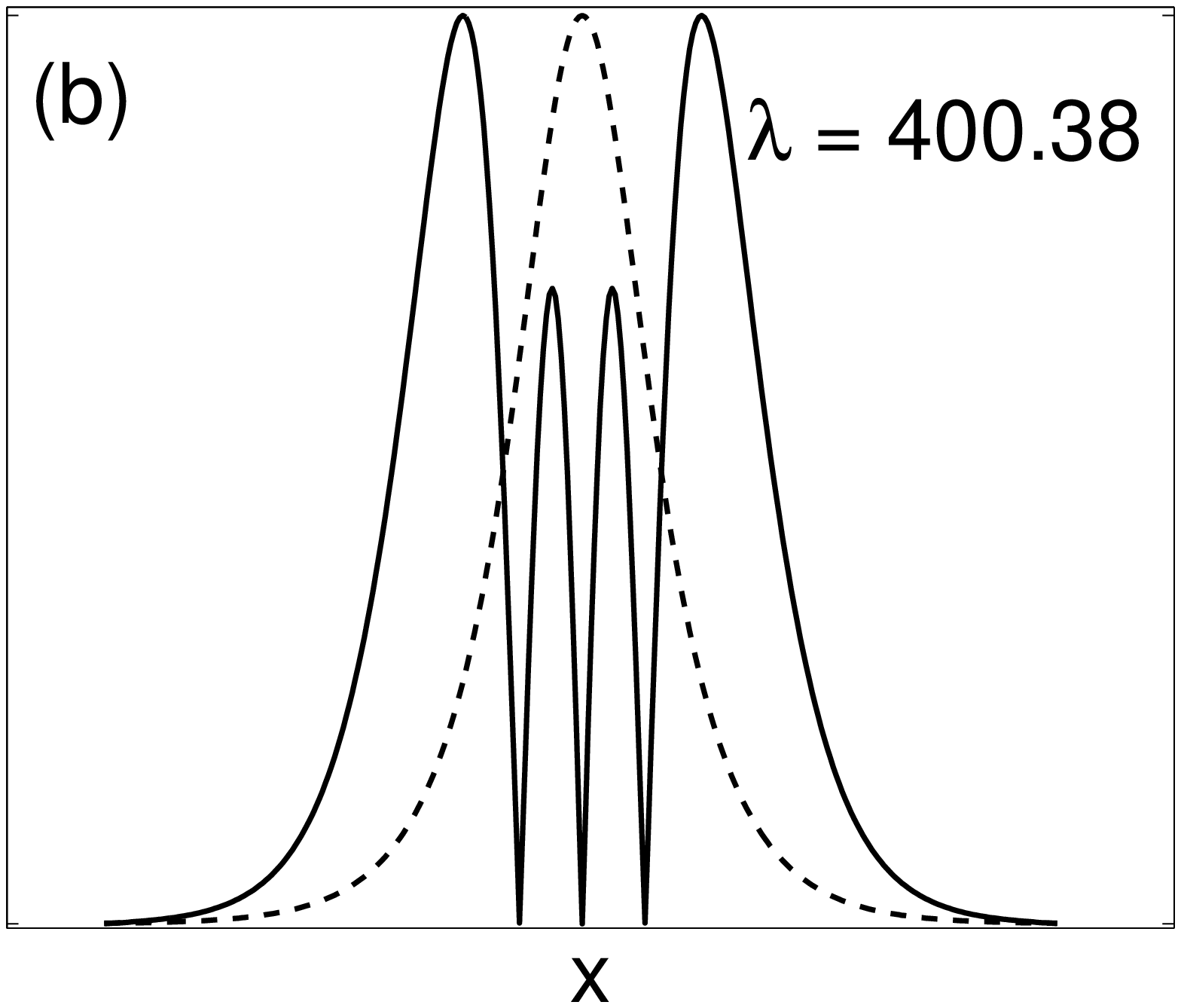}}}
 \end{minipage}
 \hspace{-0.1cm}
 \begin{minipage}{4.7cm}
   \vspace{-1.2cm}
  \rotatebox{0}{\resizebox{4.7cm}{6cm}{\includegraphics[0in,0.5in]
   [8in,10.5in]{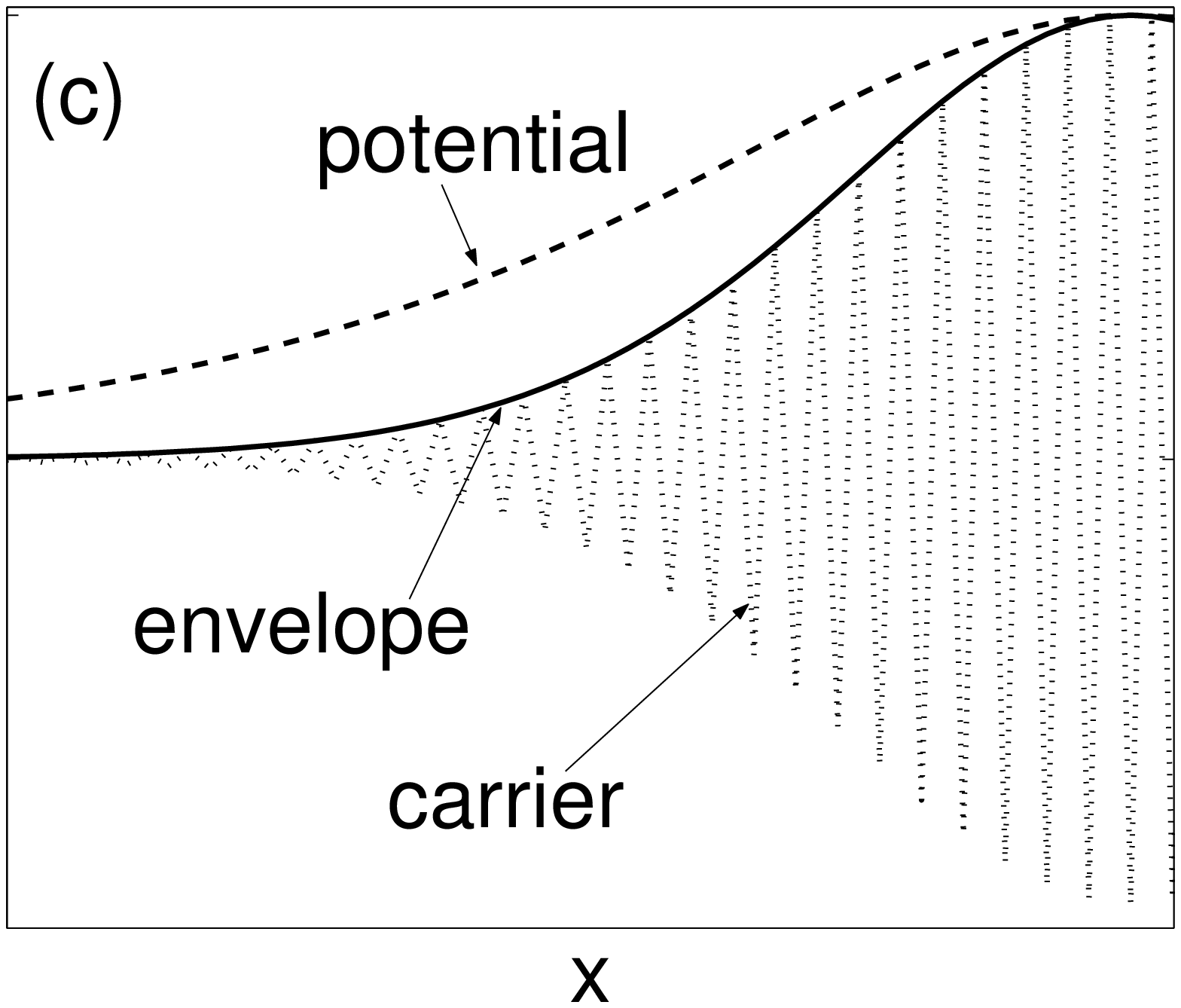}}}
  \end{minipage}
\vspace{-1.3cm}
\caption{(a,b) \ Solid lines show envelopes 
of the first and fourth highest-frequency
modes obtained by the finite-difference approximation of \eqref{e_01} with 
$V=3\,\sech(0.5x)$. Dashed line show the potential $V(x)$. The corresponding
eigenvalues are shown in each panel. \ 
(c) \ Part of panel (a), magnified and displaying the highest-frequency carrier.
}
\label{fig_1}
\end{figure}

As Fig.~1(c) illustrates, one can take \ $\psi(x_n)=\exp[i\pi n]\,\phi(x_n)$. Here the factor
\ $\exp[i (\pi/h) x_n]= \exp[i\pi n]$ \ accounts for the highest-frequency carrier,
while $\phi(x_n)$ is assumed to vary on the $x$-scale of order one. Substituting this ansatz
into \eqref{e_01} where the second derivative is approximated by \eqref{e_03}, one finds:
\be
\frac4{h^2}\phi(x) + \frac{d^2\phi(x)}{dx^2} + V(x)\phi(x) +O(h^2) = \lambda \phi.
\label{e_04}
\ee
Here $x\equiv x_n$, and we have used the Taylor expansion \ $\phi(x_{n\pm 1}) = 
\phi(x) \pm h\phi'(x) + (h^2/2)\phi''(x) + O(h^3)$ \ for the smooth envelope $\phi(x)$.
Neglecting the $O(h^2)$-term in \eqref{e_04}, one sees that that equation becomes
\be
\phi'' + (V-\D\lambda)\phi =0, \qquad \D\lambda = \lambda - (4/h^2).
\label{e_05}
\ee

Based on \eqref{e_05}, one can make the following conclusions about the appearance
of the envelopes of the 
high-frequency modes of \eqref{e_01} obtained by a finite-difference approximation.
When the potential in the original Eq.~\eqref{e_01} is repulsive, 
$V(x)>0$ (or, more generally, $\int_{-\infty}^{\infty}V(x)\,dx >0$), 
the envelope of the mode with the highest eigenvalue $\lambda$ is the ground state
of \eqref{e_05}. For smaller $\lambda$'s, one obtains consecutive excited states of
\eqref{e_05}. This is confirmed by Fig.~1(a,b). The taller and/or wider the
potential, the more high-frequency modes with localized envelopes there exist.
The non-localized envelopes, corresponding to $\D\lambda <0$ is \eqref{e_05}, 
still have spatial features on the scale of order one (see Fig.~2(a)). 
For an attractive potential in \eqref{e_01},
$V(x)<0$, there exist no localized solutions of \eqref{e_05}. In that case,
the envelopes of all high-frequency modes are not localized. Such envelopes of the
first and fifth highest-frequency modes for $V(x)=-3\,\sech(0.5x)$ are
shown in Fig.~2(b,c).

\begin{figure}[h!]
\hspace*{0.3cm}
\begin{minipage}{4.7cm}
 \vspace{-1.2cm}
\rotatebox{0}{\resizebox{4.7cm}{6cm}{\includegraphics[0in,0.5in]
 [8in,10.5in]{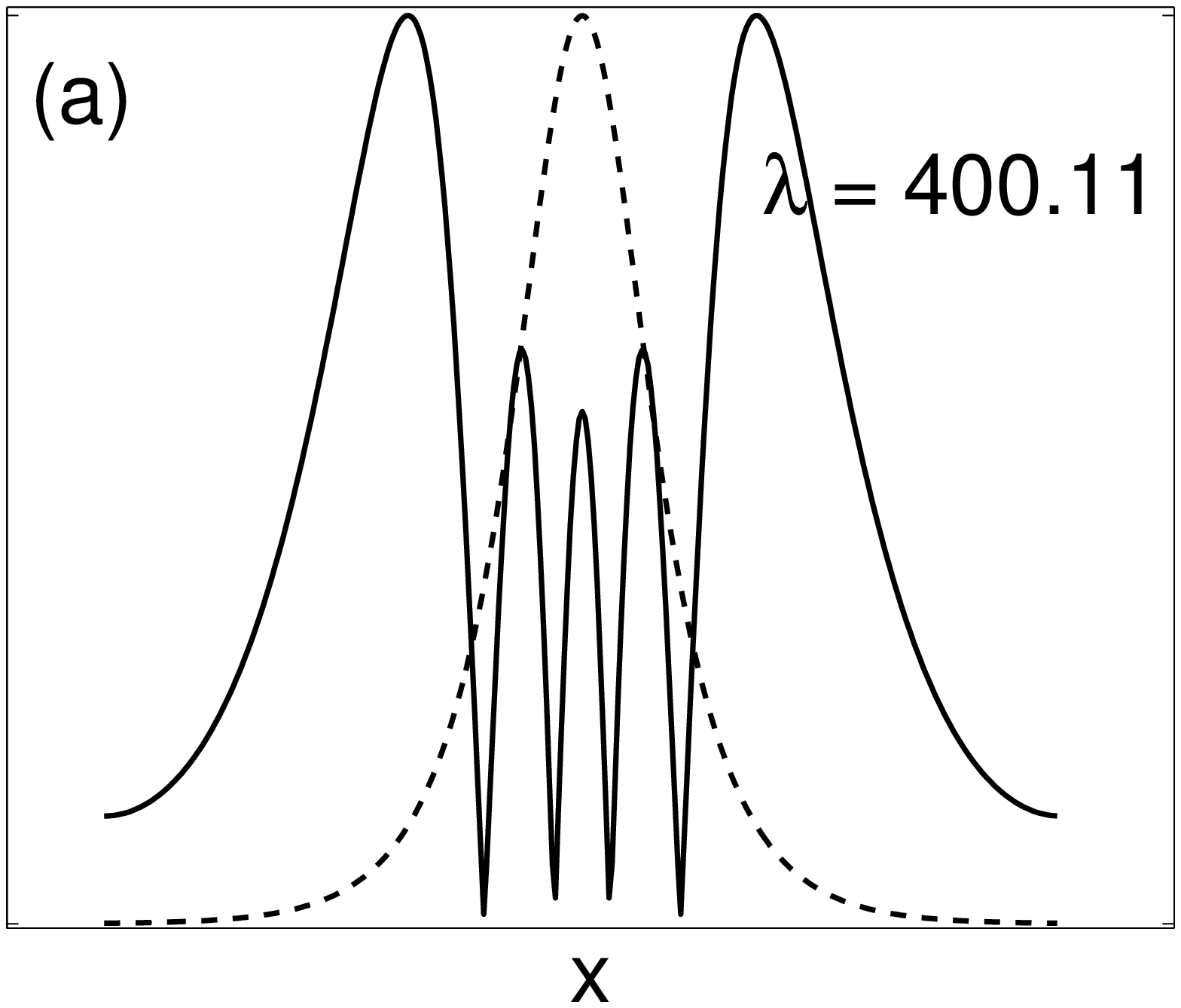}}}
 \end{minipage}
 \hspace{-0.1cm}
 \begin{minipage}{4.7cm}
 \vspace{-1.2cm}
 \rotatebox{0}{\resizebox{4.7cm}{6cm}{\includegraphics[0in,0.5in]
  [8in,10.5in]{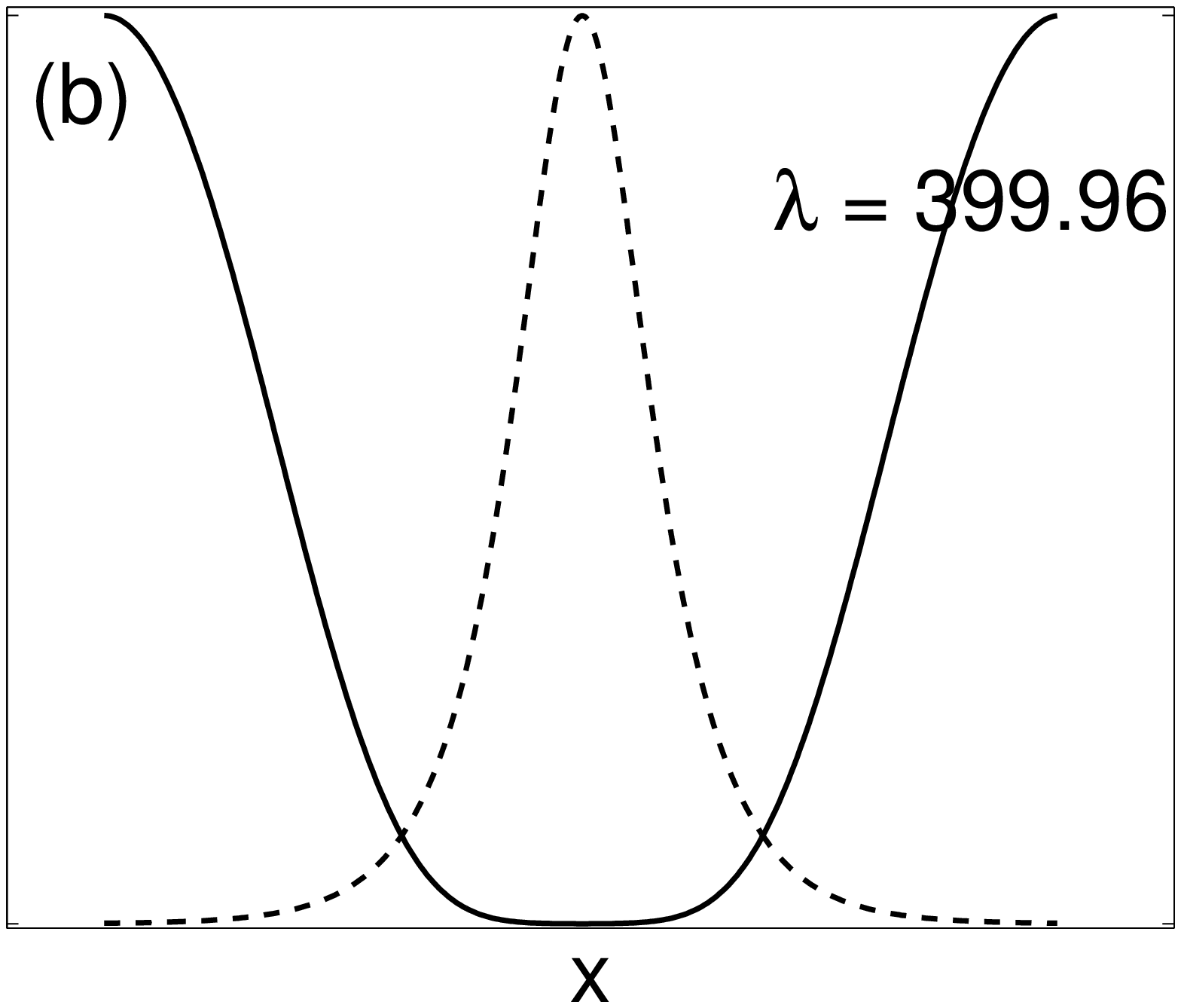}}}
 \end{minipage}
 \hspace{-0.1cm}
 \begin{minipage}{4.7cm}
   \vspace{-1.2cm}
  \rotatebox{0}{\resizebox{4.7cm}{6cm}{\includegraphics[0in,0.5in]
   [8in,10.5in]{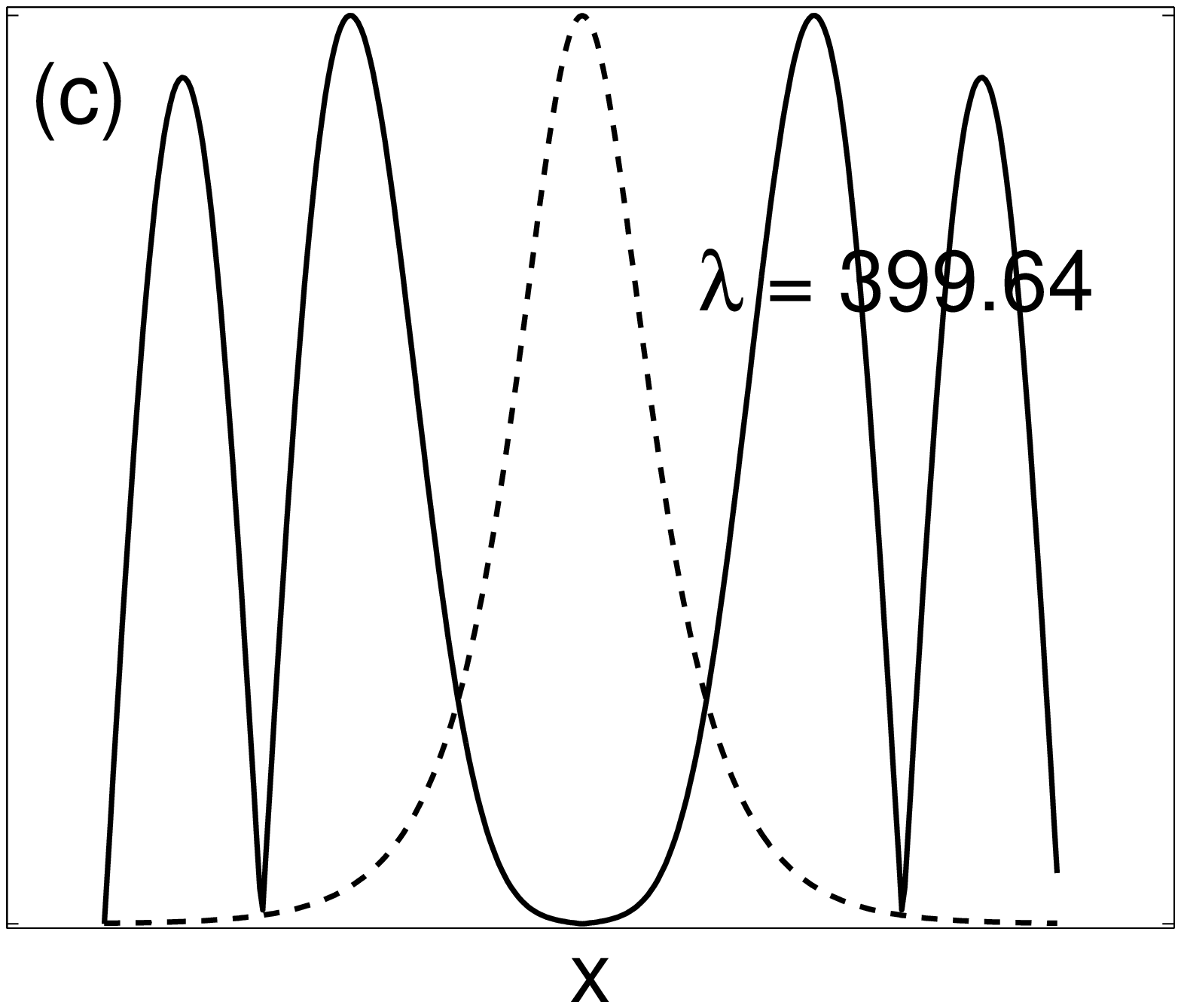}}}
  \end{minipage}
\vspace{-1.3cm}
\caption{As in Fig.~1, solid and dashed lines show the moduli of the mode envelope
and of the potential. \ (a) \ The fifth (non-localized) highest-frequency
mode obtained by the finite-difference approximation of \eqref{e_01} with 
$V=3\,\sech(0.5x)$.  \ 
(b,c) \ First and fourth highest-frequency modes (non-localized) for the
potential \ $V=-3\,\sech(0.5x)$.
}
\label{fig_2}
\end{figure}


\end{document}